\documentclass[preprint,aps,12pt,showpacs,nofootinbib,tightenlines]{revtex4}
\usepackage{amsmath}
\usepackage{amssymb}
\usepackage{CJK}
\usepackage{epsfig}
\usepackage{graphicx}
\usepackage{subfigure}
\usepackage{slashed}
\usepackage{multirow}
\newcommand{\met}{\not\!\!\!E_{T}}
\begin{document}
\def\pslash{\rlap{\hspace{0.02cm}/}{p}}
\def\eslash{\rlap{\hspace{0.02cm}/}{e}}
\title {Top partner production at $e^{+}e^{-}$ collider in the littlest Higgs Model with
T-parity}
\author{Haiyan Wang$^{1}$}
\author{Bingfang Yang$^{1,2}$}\email{yangbingfang@htu.edu.cn}

\affiliation{$^1$ School of Materials Science and Engineering, Henan
Polytechnic University, Jiaozuo 454000, China\\
$^2$College of Physics and Materials Science, Henan Normal
University, Xinxiang 453007, China
   \vspace*{1.5cm}  }

\begin{abstract}

In the framework of the littlest Higgs Model with T-parity, we
discuss the top partner production at future $e^{+}e^{-}$ collider.
We calculate the cross sections of the top partner production
processes and associated production processes of Higgs and top
partner under current constraints. Then, we investigate the
observability of the T-odd top partner pair production through the
process $e^{+}e^{-}\rightarrow T_{-}\bar{T}_{-}\rightarrow
t\bar{t}A_{H}A_{H}$ in the $t\bar{t}$ di-lepton channel for two
T-odd top partner mass $m_{T_{-}}$=603(708) GeV at $\sqrt{s}$=1.5
TeV. We analyze the signal significance depending on the integrated
luminosity and find this signal is promising at the future high
energy $e^{+}e^{-}$ collider.

\end{abstract}
\pacs{14.65.Ha,13.66.Hk,12.60.-i} \maketitle
\section{ Introduction}
\noindent The discovery of the Higgs boson at the Large Hadron
Collider (LHC) \cite{LHC-higgs} is a great step towards
understanding the electroweak symmetry breaking (EWSB) mechanism.
However, the little hierarchy problem\cite{hierarchy}, which is
essentially from quadratically divergent corrections to the Higgs
mass parameter, still exists. In the past, various new physics
models have been proposed to solve this problem, and the littlest
Higgs Model with T-parity (LHT) \cite{LHT} is one of the most
promising candidates.

In the LHT model, the Higgs boson is constructed as a
pseudo-Nambu-Goldstone particle of the broken global symmetry. The
quadratic divergence contributions to Higgs boson mass from the SM
top quark loop, gauge boson loops and the Higgs self-energy are
cancelled by the corresponding T-parity partners, respectively.
Among the partners, the top partner is the most important one since
it is responsible for cancelling the largest quadratically divergent
correction to the Higgs mass induced by the top quark.

Recently, the ATLAS and CMS collaborations have performed the
searches for the vector-like top partner through the pair or single
production with three final states $bW$, $tZ$ and $tH$, and have
excluded the top partner with the mass less than about 700 GeV
\cite{tp-LHC}. Besides, a search has been performed in pair-produced
exotic top partners, each decay to an on-shell top (or antitop)
quark and a long-lived undetected neutral particle\cite{tm-LHC}.
Apart from direct searches, the indirect searches for the top
partners through their contributions to the electroweak precision
observables (EWPOs) \cite{lht-ewpos}, $Z$-pole observables
\cite{lht-rb} and the flavor physics \cite{lh-flavor} have been
extensively investigated. The recent research\cite{directbound} shows that the direct bounds on the heavy vector-like top quarks have been
stronger than the indirect constraints. The null results of the top partners, in
conjunction with the EWPOs and the recent Higgs data, have tightly
constrained the parameter space of the LHT
model\cite{lht-fit1,lht-fit2}.

Compared to the hadron colliders, $e^{+}e^{-}$ linear colliders may
provide cleaner environments to study productions and decays of
various particles. Some design schemes have been put forward, such
as the International Linear Collider(ILC) \cite{ILC}and the Compact
Linear Collider(CLIC) \cite{CLIC}, they can run at the center of
mass (c.m.) energy ranged from 500 GeV to 3000 GeV, which enables us
to perform precision measurements of the top partner above the
threshold. In addition, the polarization of the initial beams at
$e^{+}e^{-}$ linear colliders will be useful to study the properties
of the top partner. Some relevant works have been widely studied in
various extensions of the Standard Model (SM)\cite{NP-top},
including the Little Higgs model\cite{LH-top}. However, the works in
Little Higgs model mostly were performed many years ago and before
the discovery of the Higgs boson, so it is necessary to revisit this
topic. Moreover, the different final states are analyzed in this
work.

The paper is organized as follows. In Sec.II we review the top
partner in the LHT model. In Sec.III we calculate top partner
production cross sections. In Sec.IV we investigate signal and
discovery potentiality of the top partner production at $e^{+}e^{-}$
collider. Finally, we draw our conclusions in Sec.V.

\section{Top partner in the LHT model}\label{section2}

The LHT model is a non-linear $\sigma$ model based on the coset
space $SU(5)/SO(5)$\cite{littlehiggs}. The global group $SU(5)$ is
spontaneously broken into $SO(5)$ at the scale $f\sim \mathcal
O$(TeV) by the vacuum expectation value (VEV) of the $\Sigma$ field,
which is given by
\begin{eqnarray}
\Sigma_0=\langle\Sigma\rangle=
\begin{pmatrix}
{\bf 0}_{2\times2} & 0 & {\bf 1}_{2\times2} \\
0 & 1 &0 \\
{\bf 1}_{2\times2} & 0 & {\bf 0}_{2\times 2}
\end{pmatrix}.
\end{eqnarray}
The VEV $\Sigma_0$ also breaks the gauged subgroup $\left[ SU(2)
\times U(1) \right]^{2} $ of $SU(5)$ down to the diagonal SM
electroweak symmetry $SU(2)_L \times U(1)_Y$. After the symmetry
breaking, there arise 4 new heavy gauge bosons
$W_{H}^{\pm},Z_{H},A_{H}$ whose masses given at $\mathcal
O(v^{2}/f^{2})$ by
\begin {equation}
M_{W_{H}}=M_{Z_{H}}=gf(1-\frac{v^{2}}{8f^{2}}),~~M_{A_{H}}=\frac{g'f}{\sqrt{5}}
(1-\frac{5v^{2}}{8f^{2}})
\end {equation}
with $g$ and $g'$ being the SM $SU(2)_L$ and $U(1)_Y$ gauge
couplings, respectively. The heavy photon $A_{H}$ is the lightest
$T$-odd particle and can serve as a candidate for dark matter. In
order to match the SM prediction for the gauge boson masses, the VEV
$v$ needs to be redefined as
\begin{equation}
v = \frac{f}{\sqrt{2}} \arccos{\left( 1 -
\frac{v_\textrm{SM}^2}{f^2} \right)} \simeq v_\textrm{SM} \left( 1 +
\frac{1}{12} \frac{v_\textrm{SM}^2}{f^2} \right) ,
\end{equation}
where $v_{SM}$ = 246 GeV.

In the fermion sector, the implementation of T-parity requires the
existence of mirror partners for each original fermion. In order to
do this, two fermion $SU(2)$ doublets $q_1$ and $q_2$ are introduced
and $T$-parity interchanges these two doublets. A $T$-even
combination of these doublets is taken as the SM fermion doublet and
the $T$-odd combination is its $T$-parity partner. The doublets
$q_1$ and $q_2$ are embedded into incomplete $SU(5)$ multiplets
$\Psi_1$ and $\Psi_2$ as $\Psi_1 = (q_1, 0, 0_2)^T$ and $ди2 = (0_2,
0, q_2)^T$, where $0_2 = (0, 0)^T$. To give the additional fermions
masses, an $SO(5)$ multiplet $\Psi_c$ is also introduced as
$\Psi_c=(q_c,\chi_c,\tilde{q}_c)^T$, whose transformation under the
$SU(5)$ is non-linear: $\Psi_c \to U\Psi_c$, where $U$ is the
unbroken $SO(5)$ rotation in a non-linear representation of the
$SU(5)$. The components of the latter $\Psi_c$ multiplet are the
so-called mirror fermions. Then, one can write down the following
Yukawa-type interaction to give masses of the mirror fermions
\begin{eqnarray}
\mathcal{L}_{\textrm{mirror}}=-\kappa_{ij}f\left(\bar\Psi_2^i\xi +
  \bar\Psi_1^i\Sigma_0\Omega\xi^\dagger\Omega\right)\Psi_c^j+h.c.
\end{eqnarray}
where $i, j=1,2,3$ are the generation indices. The masses of the
mirror quarks $u_{H}^{i}, d_{H}^{i}$ and mirror leptons $l_{H}^{i},
\nu_{H}^{i}$ up to $\mathcal O(v^{2}/f^{2})$ are given by
\begin{eqnarray}
m_{d_{H}^{i}}&=&\sqrt{2}\kappa_if, ~~m_{u_{H}^{i}}=
m_{d_{H}^{i}}(1-\frac{v^2}{8f^2}),\\
m_{l_{H}^{i}}&=&\sqrt{2}\kappa_if, ~~m_{\nu_{H}^{i}}=
m_{l_{H}^{i}}(1-\frac{v^2}{8f^2}),
\end{eqnarray}
where $\kappa_i$ are the diagonalized Yukawa couplings.

In the top quark sector, two singlet fields $T_{L_1}$ and $T_{L_2}$
(and their right-handed counterparts) are introduced to cancel the
large radiative correction to the Higgs mass induced by the top
quark. Both fields are embedded together with the $q_1$ and $q_2$
doublets into the $SU(5)$ multiplets: $\Psi_{1,t} = (q_1, T_{L_1},
0_2)^T$ and $\Psi_{2,t} = (0_2, T_{L_2}, q_2)^T$. The $T$-even
combination of $q_i$ is the SM fermion doublet and the other $T$-odd
combination is its $T$-parity partner. Then, the $T$-parity
invariant Yukawa Lagrangian for the top sector can be written down
as follow:
\begin{eqnarray}
{\cal L}_t &=& - \frac{\lambda_1 f}{2\sqrt{2}}\epsilon_{ijk}\epsilon_{xy} \left[(\bar{\Psi}_{1,t})_i \Sigma_{jx} \Sigma_{ky} - (\bar{\Psi}_{2,t} \Sigma_0)_i \Sigma^{'}_{jx} \Sigma^{'}_{ky} \right] t^{'}_R \nonumber \\
&& -\lambda_2 f (\bar{T}_{L_1} T_{R_1} + \bar{T}_{L_2} T_{R_2}) +
~{\rm h.c.}
\end{eqnarray}
where $\epsilon_{ijk}$ and $\epsilon_{xy}$ are the antisymmetric
tensors with $i,j,k=1,2, 3$ and $x, y=4, 5$,
$\Sigma^{'}=\langle\Sigma\rangle\Omega\Sigma^\dagger\Omega\langle\Sigma\rangle$
is the image of $\Sigma$ under $T$-parity, $\lambda_1$ and
$\lambda_2$ are two dimensionless top quark Yukawa couplings. Under
$T$-parity, these fields transform as: $T_{L_1} \leftrightarrow -
T_{L_2}$, $T_{R_1} \leftrightarrow -T_{R_2}$, $t^{'}_R \to t^{'}_R$.
The above Lagrangian contains the following mass terms:
\begin{eqnarray}
{\cal L}_t \supset - \lambda_1 f \left( \frac{s_\Sigma}{\sqrt{2}}
\bar{t}_{L_+} t^{'}_R + \frac{1+c_\Sigma}{2}\bar{T}^{'}_{L_+}t^{'}_R
\right) -\lambda_2 f (\bar{T}^{'}_{L_+}T^{'}_{R_+} +
\bar{T}^{'}_{L_-}T^{'}_{R_-}) + ~{\rm h.c.}
\end{eqnarray}
where $c_\Sigma = \cos(\sqrt{2}h/f)$ and $s_\Sigma =
\sin(\sqrt{2}h/f)$. The $T$-parity eigenstates have been defined as
$t_{L_+}=(t_{L_1}-t_{R_1})/\sqrt{2}$, $T^{'}_{L_\pm}=(T_{L_1} \mp
T_{L_2})/\sqrt{2}$ and $T^{'}_{R_\pm}=(T_{R_1} \mp
T_{R_2})/\sqrt{2}$. Note that T-odd Dirac fermion $T_{-}\equiv
(T^{'}_{L_-},T^{'}_{R_-})$ does not have the tree level Higgs boson
interaction, and thus it does not contribute to the Higgs mass at
one-loop level.

The two T-even eigenstates $(t_{L_+},t^{'}_R)$ and
$(T^{'}_{L_+},T^{'}_{R_+})$ mix with each other so that the mass
eigenstates can be defined as
\begin{eqnarray}
t_L &=& \cos\beta \,t_{L_+} - \sin\beta \,T^{'}_{L_+}, \quad T_{L_+} = \sin\beta \,t_{L_+} +\cos\beta \,T^{'}_{L_+},\nonumber \\
t_R &=& \cos\alpha \,t^{'}_R - \sin\alpha \,T^{'}_{R_+}, ~\quad
T_{R_+} = \sin\alpha \,t^{'}_R + \cos\alpha \,T^{'}_{R_+},\
\label{combination}
\end{eqnarray}
where the mixing angles $\alpha$ and $\beta$ can be defined by the
dimensionless ratio $R=\lambda_1/\lambda_2$ as,
\begin{eqnarray}
\sin\alpha=\frac{R}{\sqrt{1+R^2}}, \quad
\sin\beta=\frac{R^2}{1+R^2}\frac{v}{f}.
\end{eqnarray}
The $t\equiv (t_{L}, t_{R})$ quark is identified with the SM top
quark, and $T_+ \equiv (T_{L_+}, T_{R_+})$ is its T-even heavy
partner, which is responsible for the cancellation of the quadratic
divergence to the Higgs mass induced by the top quark loop.

The Yukawa term generates the masses of the top quark and its
partners, which are given at $\mathcal O(v^{2}/f^{2})$ by
\begin{eqnarray}
&&m_t=\frac{\lambda_2 v R}{\sqrt{1+R^2}} \left[ 1 + \frac{v^2}{f^2}
\left( -\frac{1}{3} + \frac{1}{2} \frac{R^2}{(1+R^2)^2} \right)\right]\nonumber \\
&&m_{T_{+}}=\frac{f}{v}\frac{m_{t}(1+R^2)}{R}\left[1+\frac{v^{2}}{f^{2}}\left(\frac{1}{3}-\frac{R^2}{(1+R^2)^2})\right)\right] \nonumber \\
&&m_{T_{-}}=\frac{f}{v}\frac{m_{t}\sqrt{1+R^2}}{R}\left[1+\frac{v^{2}}{f^{2}}\left(\frac{1}{3}-\frac{1}{2}\frac{R^2}{(1+R^2)^2})\right)\right]\label{Tmass}
\end{eqnarray}
Since the $T_{+}$ mass is always larger than the T-odd top partner
$T_{-}$ mass, the $T_{+}$ can decay into $A_{H}T_{-}$ in addition to
the conventional decay modes ($Wb, tZ, tH$).

The T-invariant Lagrangians of the Yukawa interactions of the
down-type quarks and charged leptons can be constructed by two
possible ways, which are denoted as Case A and Case B,
respectively\cite{caseAB}. In the two cases, the corrections to the
Higgs couplings with the down-type quarks and charged leptons with
respect to their SM values are given at order $\mathcal{O} \left(
v_{SM}^4/f^4 \right)$ by ($d \equiv d,s,b,\ell^{\pm}_i$)
\begin{eqnarray}
    \frac{g_{h \bar{d} d}}{g_{h \bar{d} d}^{SM}} &=& 1-
        \frac{1}{4} \frac{v_{SM}^{2}}{f^{2}} + \frac{7}{32}
        \frac{v_{SM}^{4}}{f^{4}} \qquad \text{Case A} \nonumber \\
    \frac{g_{h \bar{d} d}}{g_{h \bar{d} d}^{SM}} &=& 1-
        \frac{5}{4} \frac{v_{SM}^{2}}{f^{2}} - \frac{17}{32}
        \frac{v_{SM}^{4}}{f^{4}} \qquad \text{Case B}
    \label{dcoupling}
\end{eqnarray}

\section{Top partner production in $e^{+}e^{-}$ collision}

\noindent
\begin{figure}[htbp]
\scalebox{0.5}{\epsfig{file=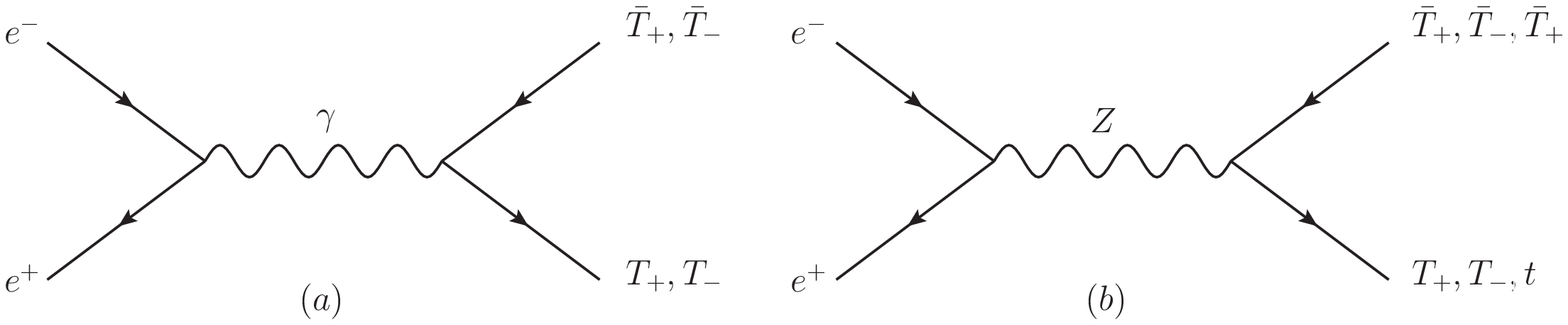}}\vspace{-0.5cm}
\caption{Feynman diagrams of the top partner production at
$e^{+}e^{-}$ collider.}\label{eett}
\end{figure}
\begin{figure}[htbp]
\scalebox{0.5}{\epsfig{file=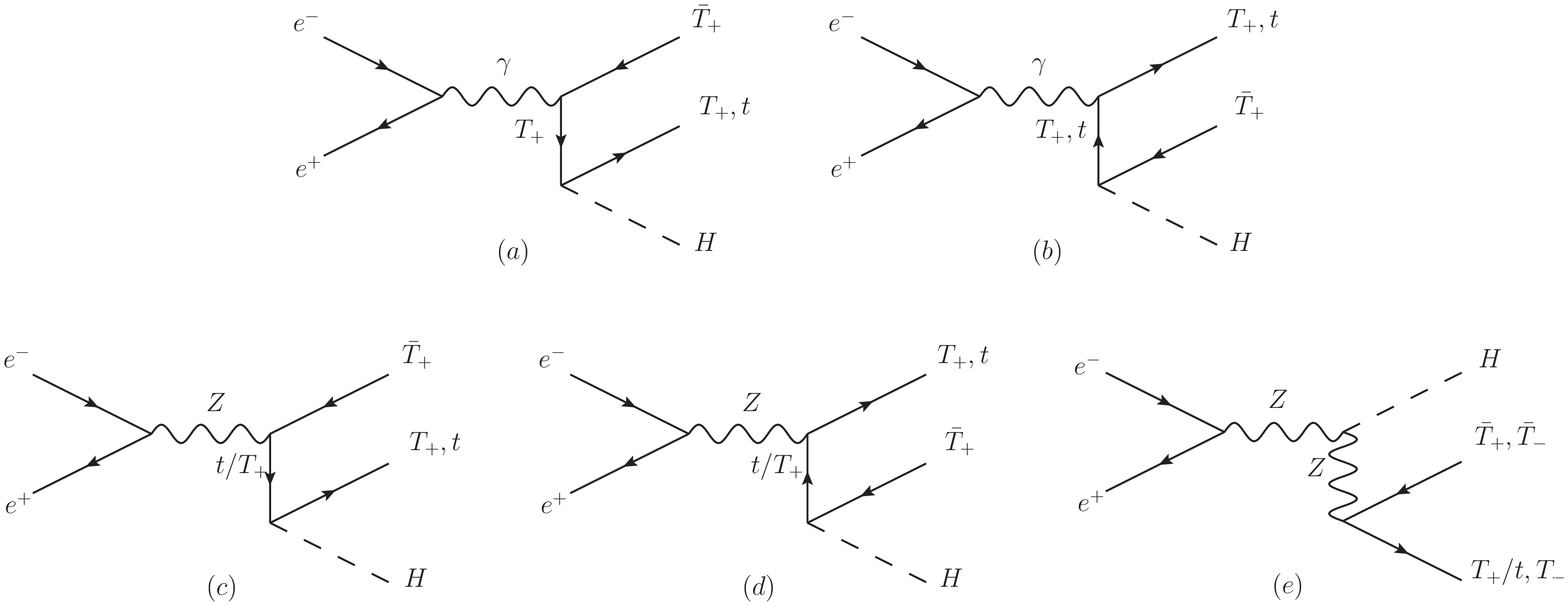}}\vspace{-0.5cm}\caption{Feynman
diagrams of the Higgs and top partner associated production at
$e^{+}e^{-}$ collider.}\label{eetth}
\end{figure}
\begin{figure}[htbp]
\begin{center}
\scalebox{0.3}{\epsfig{file=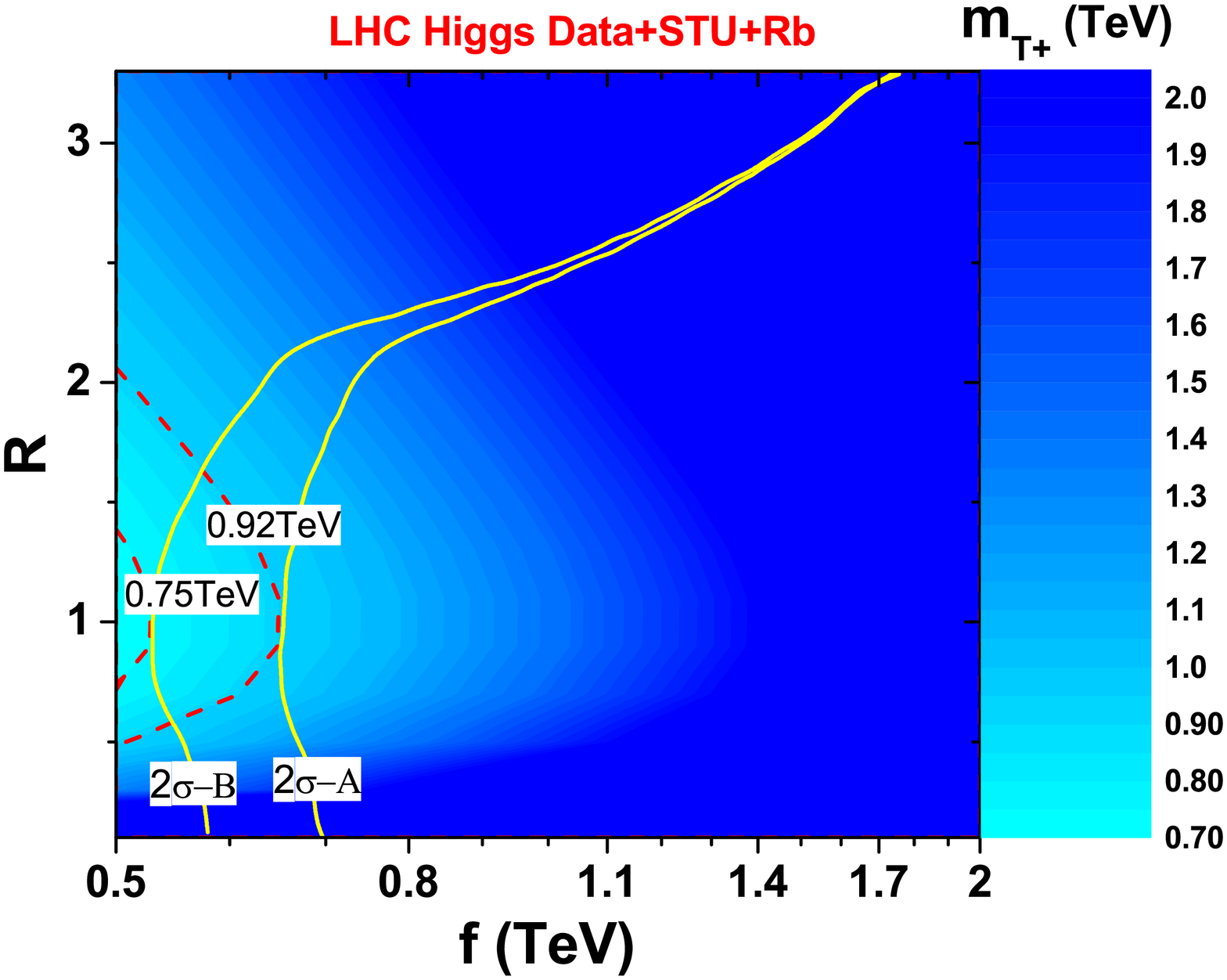}}\vspace{-0.5cm}\hspace{-0.5cm}
\scalebox{0.3}{\epsfig{file=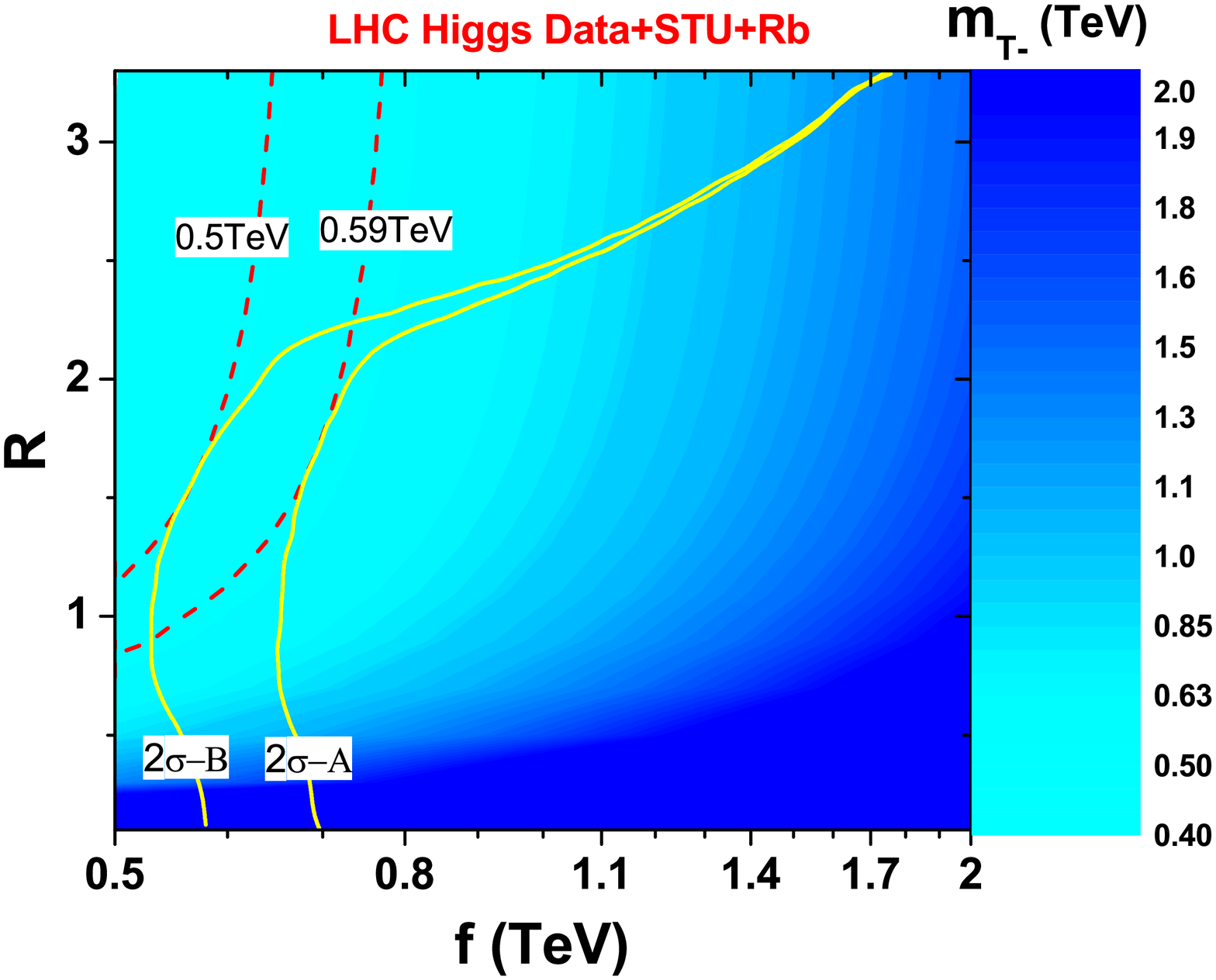}} \caption{Exclusion limits on
the top partner masses on the $R\sim f$ plane at $2\sigma$
confidence level for Case A and Case B, where the parameter $\kappa$
is marginalized over. }\label{mass}
\end{center}
\end{figure}

In the LHT model, the Feynman diagrams of top partner production are
shown in Fig.\ref{eett}, which proceeds through the $s$-channel
$\gamma$ and $Z$ exchange diagrams. These processes include T-even
top partner pair production $e^{+}e^{-}\rightarrow
T_{+}\bar{T}_{+}$, T-odd top partner pair production
$e^{+}e^{-}\rightarrow T_{-}\bar{T}_{-}$ and a T-even top partner
associating with a top quark production $e^{+}e^{-}\rightarrow
t\bar{T}_{+}$.

The Feynman diagrams of the Higgs and top partner associated
production are shown in Fig.\ref{eetth}, which has additional
diagrams mediated by the T-even top partner $T_{+}$ compared to the
process $e^{+}e^{-}\rightarrow t\bar{t}H$ in the SM. These processes
include Higgs associating with T-even top partner pair production
$e^{+}e^{-}\rightarrow T_{+}\bar{T}_{+}H$, Higgs associating with
T-odd top partner pair production $e^{+}e^{-}\rightarrow
T_{-}\bar{T}_{-}H$ and Higgs associating with a top quark and a
T-even top partner production $e^{+}e^{-}\rightarrow t\bar{T}_{+}H$.

Before calculating the top partner production cross section, we
firstly consider the constraints on the top partner mass from
current measurements. We update the constraint on the LHT parameter
in our previous works\cite{prework}, where the global fit of the
latest Higgs data, EWPOs and $R_{b}$ measurements is performed.
Thereinto, the constraints from the direct searches for Higgs data
at Tevatron \cite{CDF}\cite{D0}and LHC\cite{ATLAS}\cite{CMS} are
obtained by the package
\textsf{HiggsSignals-1.4.0}\cite{HiggsSignals}, which is linked to
the \textsf{HiggsBounds-4.2.1}\cite{HiggsBounds} library. We compute
the $\chi^{2}$ values by the method introduced in Ref.\cite{method}
and obtained the constraint on the LHT parameter space. This
constraint will lead to the exclusion limits on the top partner
masses, which is displayed on the $R\sim f$ plane for Case A and
Case B in Fig.\ref{mass} at $2\sigma$ confidence level with
$\delta\chi^{2}=8.02$. We can see that the combined constraints can
respectively exclude $m_{T_{+}}$ and $m_{T_{-}}$ up to
\begin{eqnarray}
&&m_{T_{+}}>920(750)\rm GeV \qquad \text{Case A(B)}, \\
&&m_{T_{-}}>590(500)\rm GeV \qquad \text{Case A(B)}.
\end{eqnarray}
One can notice that Case B predicts a stronger suppression for the
down-type fermion couplings to the Higgs boson, such as $Hb\bar{b}$,
which helps to enhance the branching ratios of $H\rightarrow
\gamma\gamma, WW^{\ast}, ZZ^{\ast}, \tau\tau$, so that the Case B is
favored by the experimental data\cite{LHChiggs}.

\begin{figure}[htbp]
\begin{center}
\scalebox{0.32}{\epsfig{file=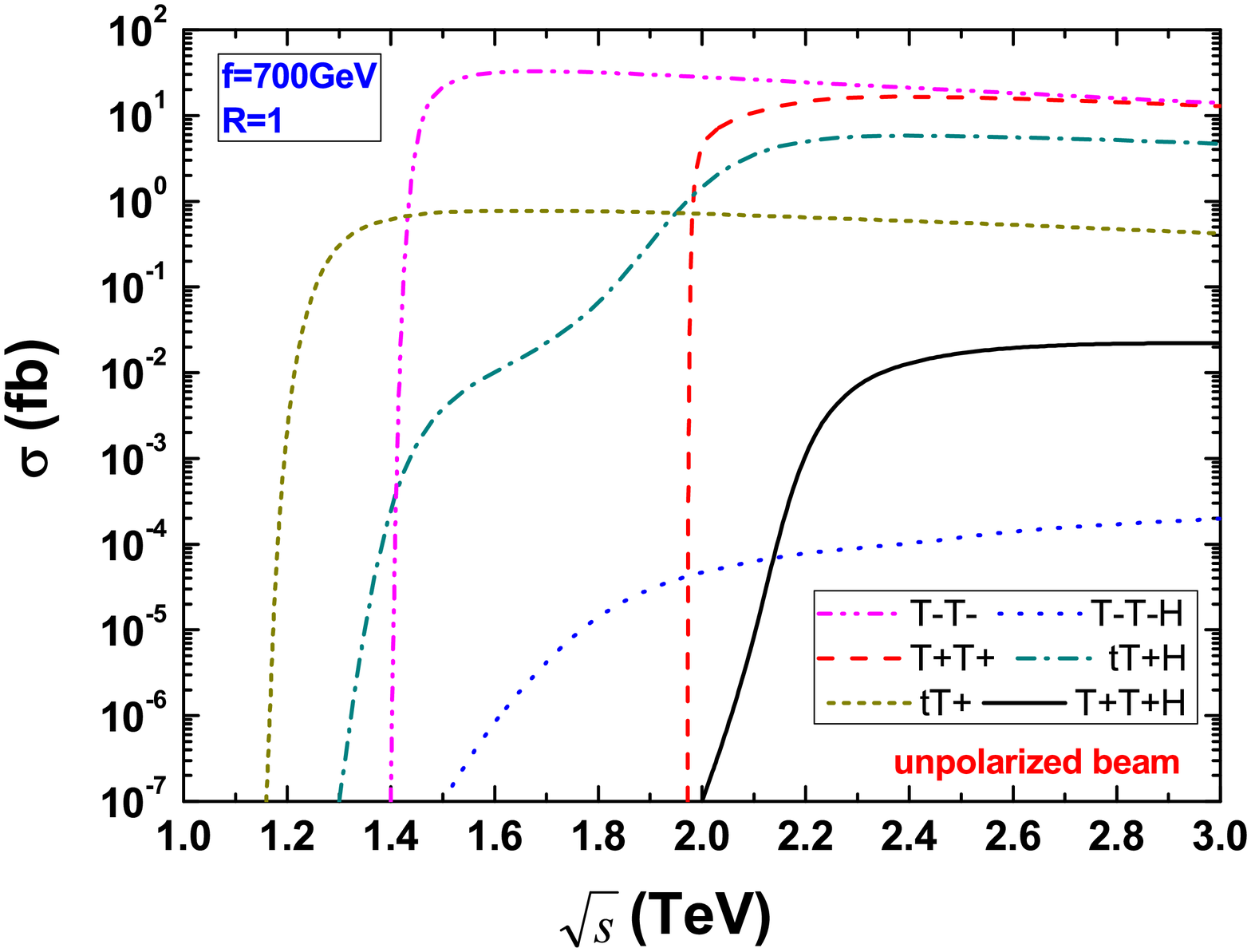}}\hspace{-0cm}
\scalebox{0.32}{\epsfig{file=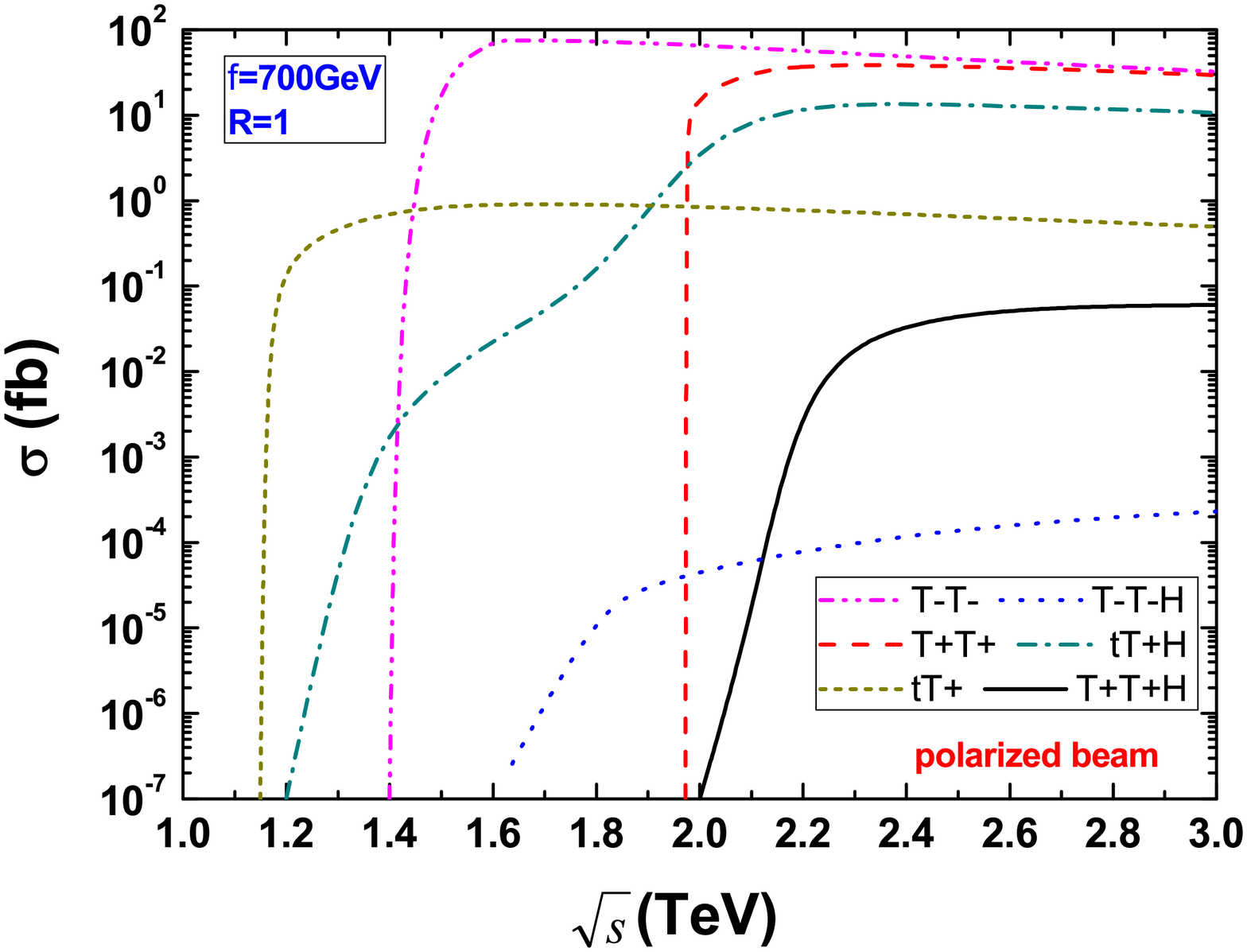}}\vspace{-0.5cm}
\caption{Top partner production cross sections as a function of
$\sqrt{s}$ for $f=700$GeV, $R=1$ in $e^{+}e^{-}$ collision with
(un)polarized beam.}\label{cross}
\end{center}
\end{figure}
In the left frame of Fig.\ref{cross}, we show the top partner
production cross sections as a function of c.m. energy $\sqrt{s}$
for $f=700$ GeV, $R=1$ (correspond to $m_{T_{+}}=986$ GeV and
$m_{T_{-}}=708$ GeV) in $e^{+}e^{-}$ collision with unpolarized
beams. The production cross sections are calculated at tree-level by
using \textsf{CalcHEP 3.6.25}\cite{calchep}, where the SM parameters
are taken as follows\cite{parameter}
\begin{eqnarray}
\nonumber
\sin^{2}\theta_{W}=0.231,~\alpha_{e}=1/128,~M_{Z}=91.1876\textrm{GeV},~m_{t}=173.5\textrm{GeV},~m_{H}=125\textrm{GeV}.
\end{eqnarray}
We can see that the top partner pair production cross sections
increase abruptly at threshold, reaches a maximum roughly 200 GeV
above threshold. Then, the production cross sections fall roughly
with the c.m. energy $\sqrt{s}$ increase due to the $s$-channel
suppression. The $T_{-}\bar{T}_{-}$ production usually has a larger
cross section than $T_{+}\bar{T}_{+}$ production since the $T_{-}$
mass is always lighter than the $T_{+}$ mass in the LHT model. The
production cross sections of the associated production of Higgs and
top partner have the similar behavior as the top partner pair
production, but usually have smaller cross sections due to smaller
phase space. The production cross section of the process
$e^{+}e^{-}\rightarrow t\bar{T}_{+}H$ reaches its maximum when the
resonance decay of the top partner $T_{+}$ emerges.

Considering the polarization of the initial electron and positron
beams, the cross section at $e^{+}e^{-}$ collider can be expressed
as\cite{polarization}
\begin{eqnarray}
\sigma&=&\frac{1}{4}\left[(1+p_{e})(1+p_{\bar{e}})\sigma_{RR}
 +(1-p_{e})(1-p_{\bar{e}})\sigma_{LL} \right. \nonumber \\
&& \left.
+(1+p_{e})(1-p_{\bar{e}})\sigma_{RL}+(1-p_{e})(1+p_{\bar{e}})\sigma_{LR}\right],
\end{eqnarray}
where $\sigma_{RL}$ is the cross section for completely right-handed
polarized $e^-$ beam ($p_{e}=+1$) and completely left-handed
polarized $e^{+}$ beam ($p_{\bar{e}}=-1$), and other cross sections
$\sigma_{RR}$, $\sigma_{LL}$ and $\sigma_{LR}$ are defined
analogously. We show the top partner production cross sections in
polarized beam with $p_{e}=0.8$ and $p_{\bar{e}}=-0.6$ in the right
frame of Fig.\ref{cross} and find that the relevant top partner
production cross sections can be enhanced by the polarized beams.

\section{SIGNAL AND DISCOVERY POTENTIALITY}
Take into account the relatively large production cross section, we
will perform the Monte Carlo simulation and explore the sensitivity
of T-odd top partner production in the following section. The T-odd
top partner $T_{-}$ has a simple decay pattern, which decays almost
100\% into the $A_{H}t$ mode. We will explore the sensitivity of
T-odd top partner pair production with unpolarized beam through the
channel,
\begin{eqnarray}
e^{+}e^{-}\rightarrow T_{-}\bar{T}_{-}\rightarrow t(\rightarrow
l^{+}\nu_{l}b)\bar{t}(\rightarrow
l^{-}\bar{\nu_{l}}\bar{b})A_{H}A_{H}\rightarrow l^{+}l^{-}+2b+\met
\end{eqnarray}
which implies that the events contain one pair of oppositely charged
leptons $l^{+}l^{-}(l=e,\mu)$ with high transverse momentum, two
high transverse momentum $b$-jets and large missing transverse
energy $\met$.

The dominant background arises from $e^{+}e^{-}\rightarrow t\bar{t}$
in the SM. Besides, the most relevant backgrounds come from
$t\bar{t}Z(\rightarrow \nu\bar{\nu})$, $W^{+}(\rightarrow
l^{+}\nu_{l})W^{-}(\rightarrow l^{-}\bar{\nu_{l}})Z(\rightarrow
b\bar{b})$ and $W^{+}(\rightarrow l^{+}\nu_{l})W^{-}(\rightarrow
l^{-}\bar{\nu_{l}})H(\rightarrow b\bar{b})$. Here, the backgrounds
$ZZZ$, $ZZH$ and $ZHH$ are neglected due to their small cross
sections. We turn off the parton-level cuts and generate the signal
and background events by using \textsf{MadGraph 5}\cite{MadGraph},
where the UFO\cite{UFO} format of the LHT model has been obtained by
FeynRules\cite{FeynRules} in Ref.\cite{lht-fit1}. We use
\textsf{MadGraph 5} to generate the process by issuing the following
commands:

generate e- e+ $>$ thodd thodd$\sim$, (thodd $>$ t ah, t $>$ l+ vl b
), (thodd$\sim$ $>$ t$\sim$ ah, t$\sim$ $>$ l- vl$\sim$ b$\sim$
)[for signal];

generate e- e+ $>$ t t$\sim$,  t $>$ l+ vl b, t$\sim$ $>$ l-
vl$\sim$ b$\sim$[for $t\bar{t}$];

generate e- e+ $>$ t t$\sim$ z,  t $>$ l+ vl b, t$\sim$ $>$ l-
vl$\sim$ b$\sim$, z $>$ vl vl$\sim$ [for $t\bar{t}Z$];

generate e- e+ $>$ w- w+ z,  w- $>$ l- vl$\sim$, w+ $>$ l+ vl, z $>$
b b$\sim$ [for $WWZ$];

generate e- e+ $>$ w- w+ h,  w- $>$ l- vl$\sim$, w+ $>$ l+ vl, h $>$
b b$\sim$ [for $WWH$].

The parton shower and hadronization are performed with
\textsf{PYTHIA}\cite{PYTHIA}, and the fast detector simulations are
performed with \textsf{Delphes}\cite{Delphes}. We use the default
card (i.e. delphes\_card\_ILD) of ILC in \textsf{Delphes 3.3.3}. The
$b$-jet tagging efficiency is taken as default value in delphes,
where it is parameterized as a function of the transverse momentum
and rapidity of the jets. When generating the parton level events,
we assume $\mu_{R}= \mu_{F}$ to be the default event-by-event value.
\textsf{FastJet}\cite{fastjet} is used to define jets via the
anti-$k_{t}$ algorithm \cite{algorithm} with distance parameter
$\Delta R =0.4$. We use \textsf{MadAnalysis 5} \cite{MadAnalysis}
for analysis, where the (mis)tagging efficiencies and fake rates are
assumed to be their default values.

\begin{figure}[htbp]
\begin{center}
\scalebox{0.4}{\epsfig{file=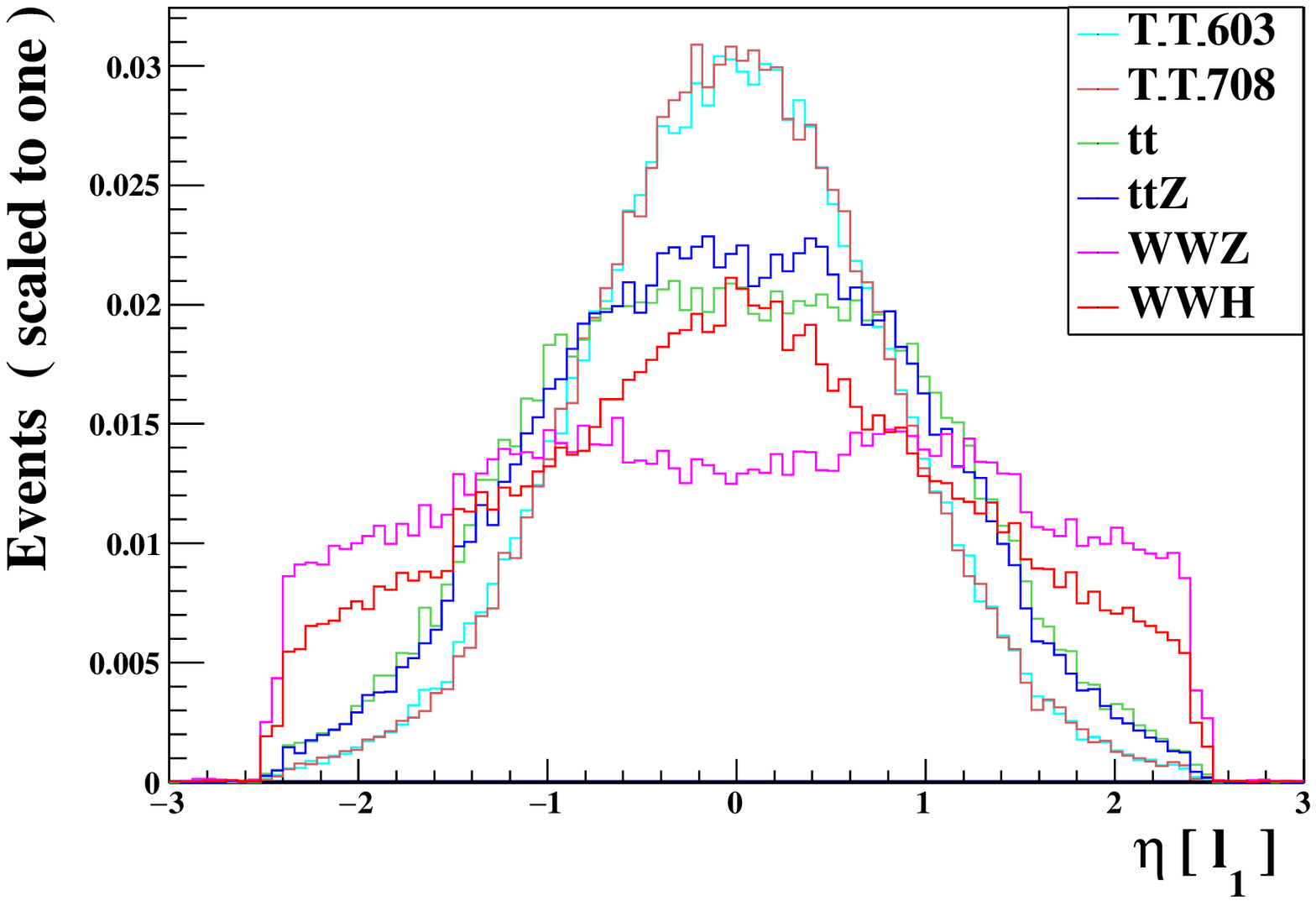}}\hspace{-0.3cm}\vspace{-0cm}
\scalebox{0.4}{\epsfig{file=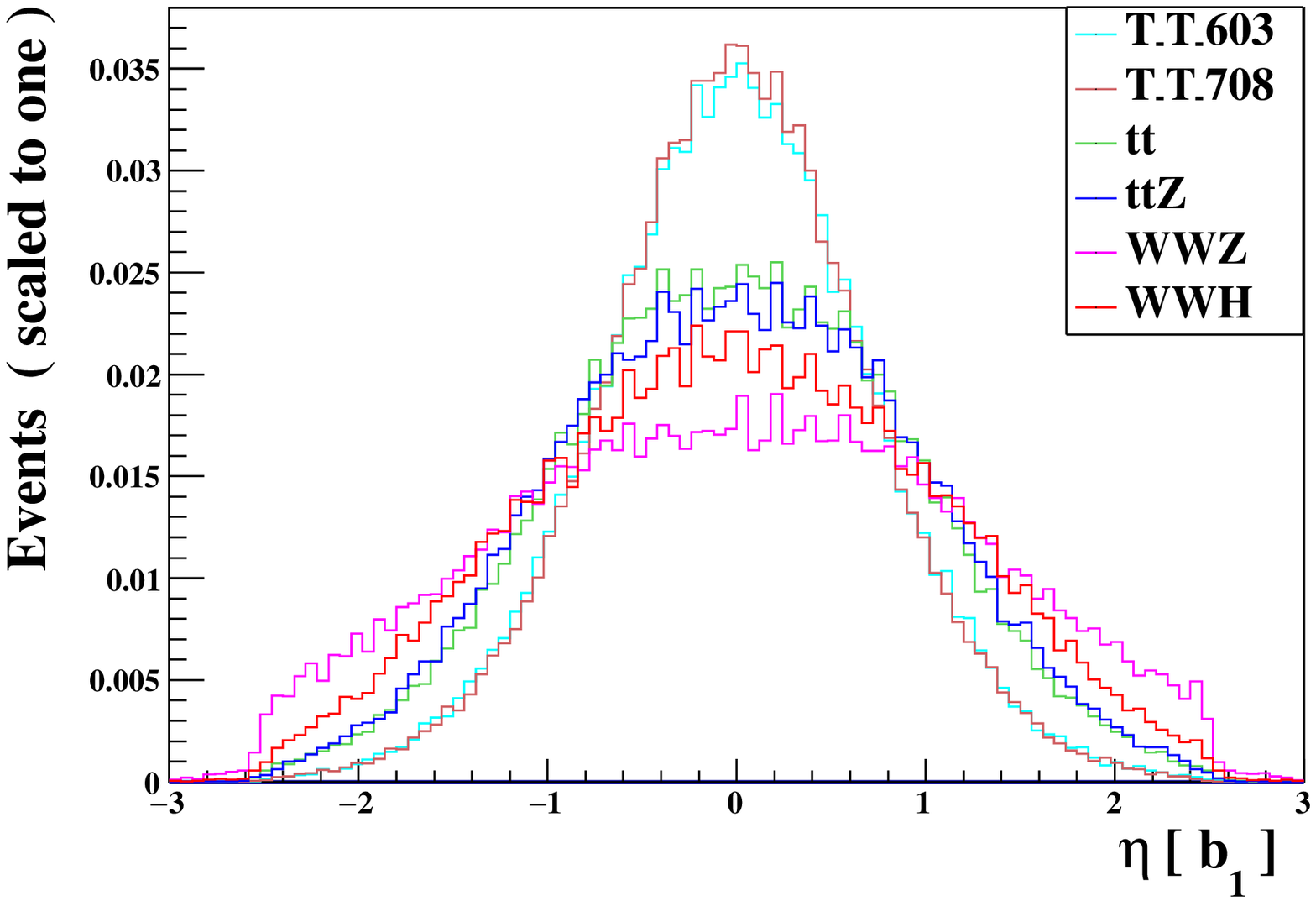}}
\scalebox{0.4}{\epsfig{file=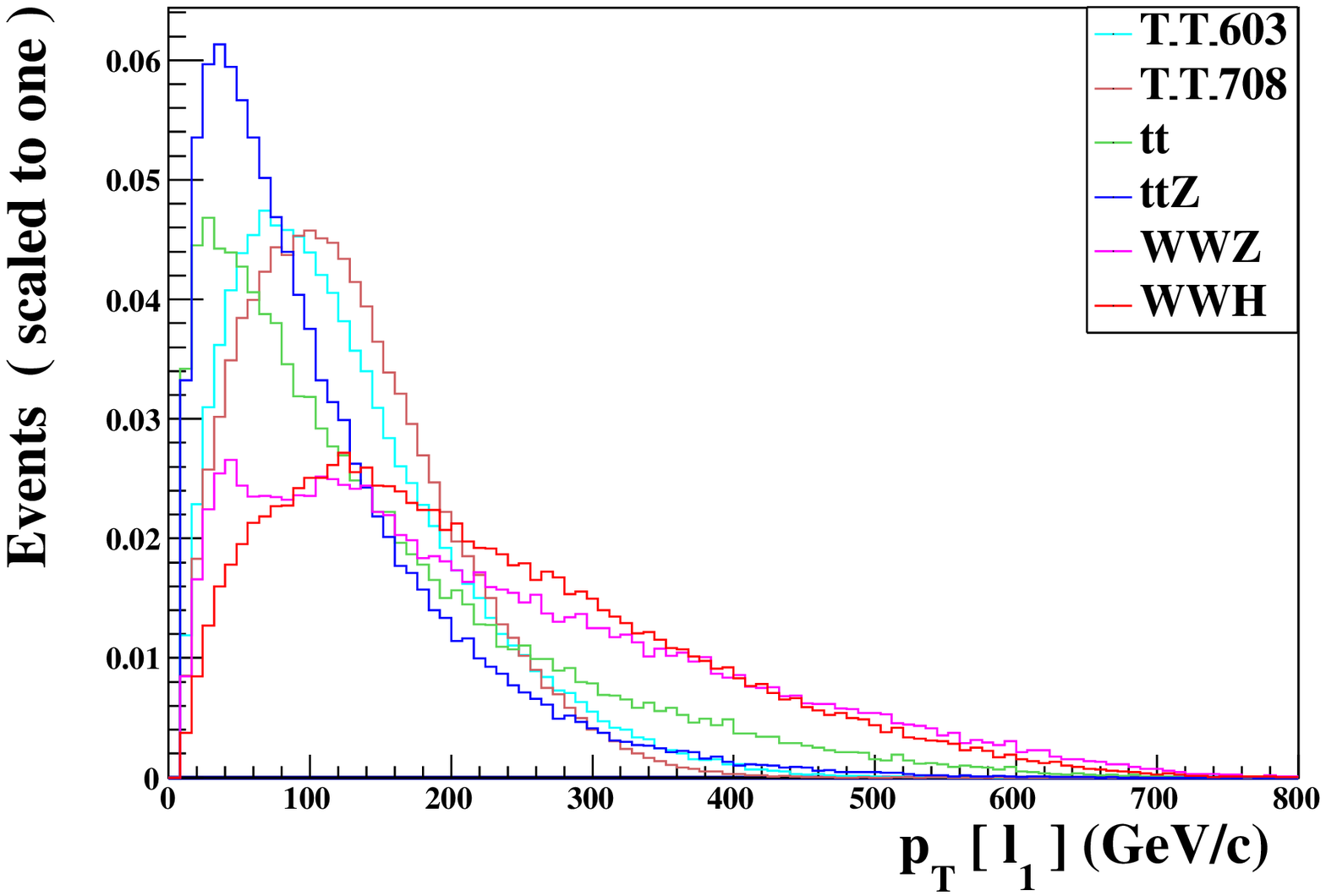}}\vspace{-0cm}\hspace{-0.3cm}
\scalebox{0.4}{\epsfig{file=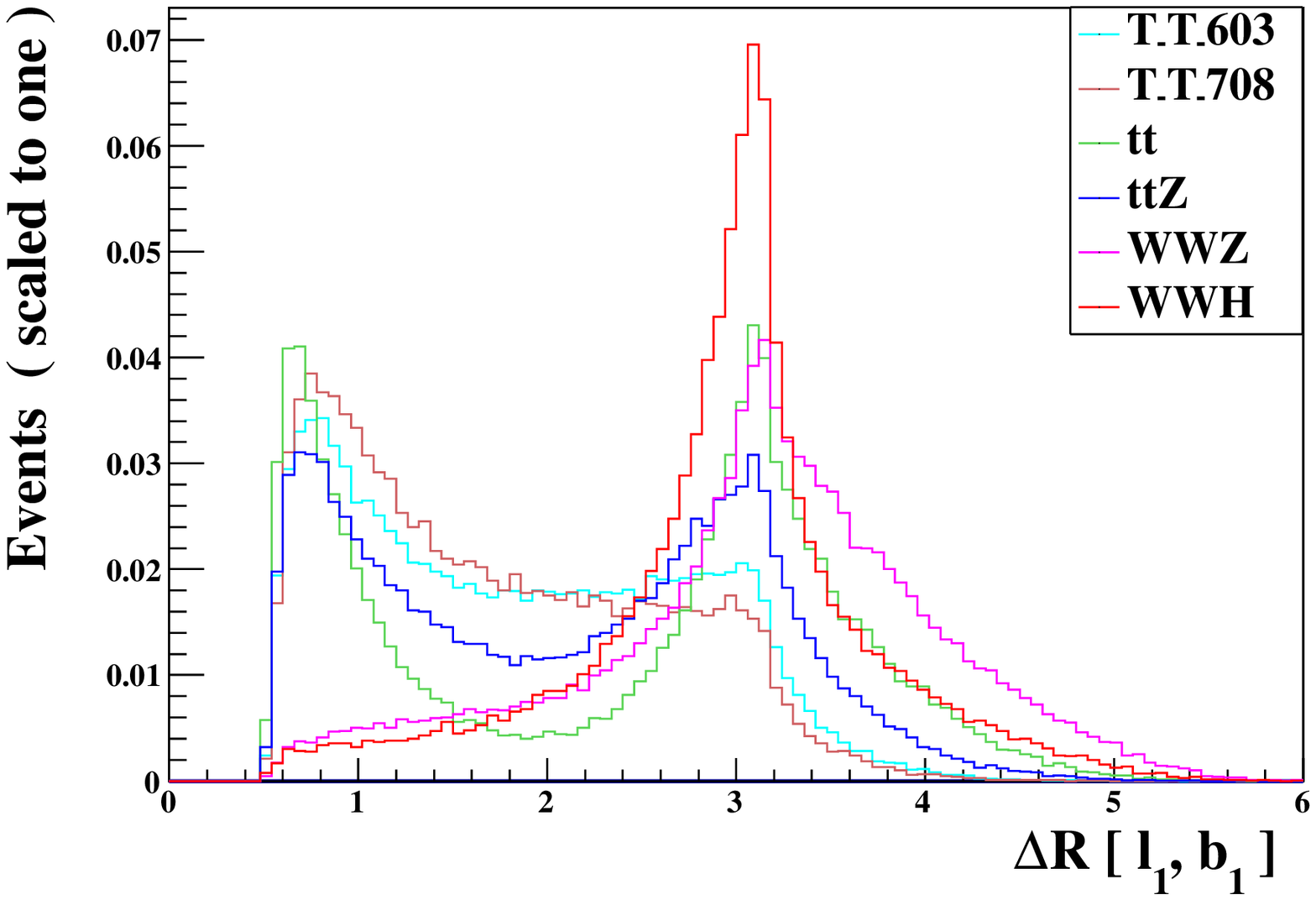}}
\scalebox{0.4}{\epsfig{file=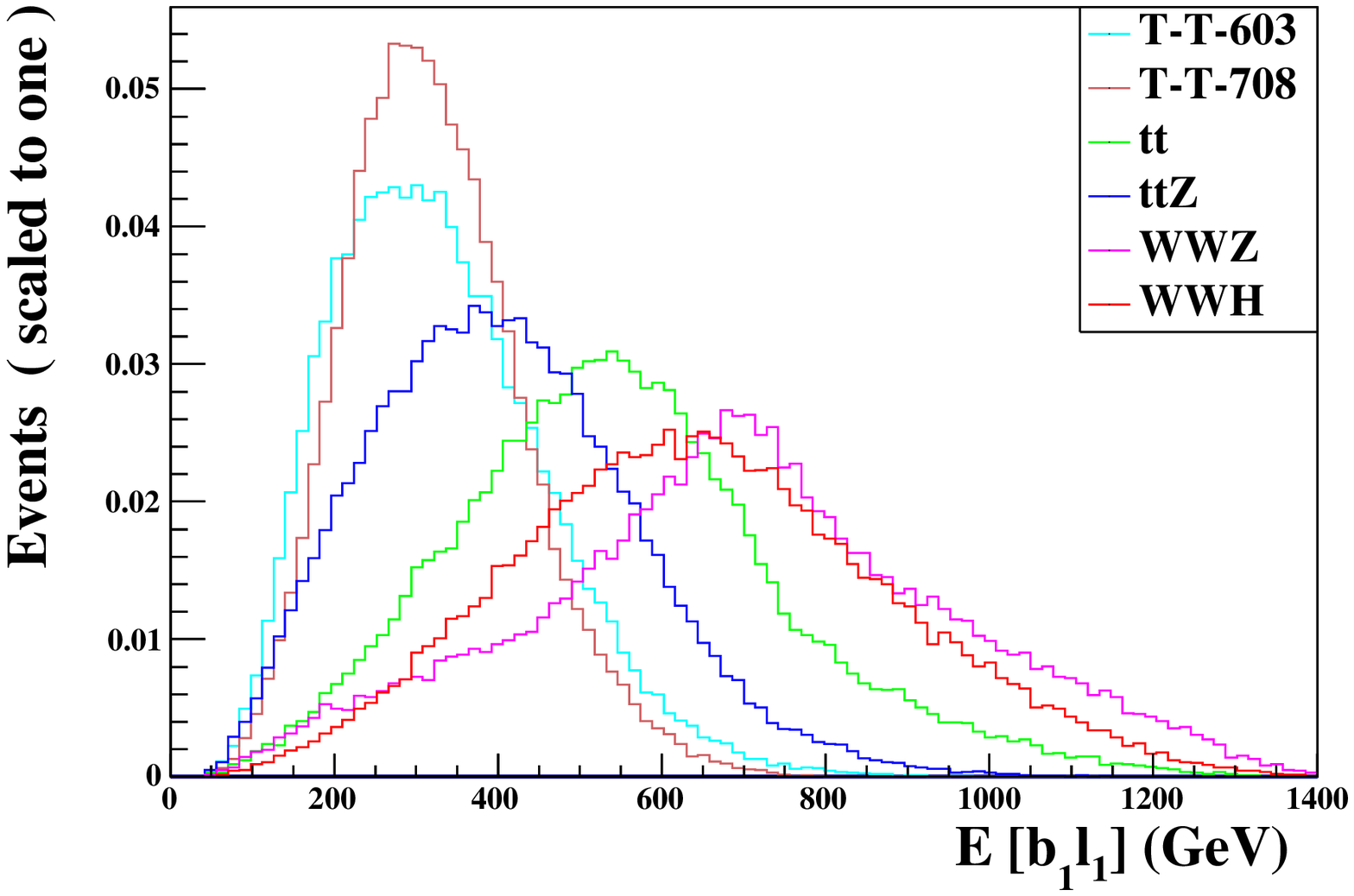}}\vspace{-0cm}\hspace{-0.3cm}
\scalebox{0.4}{\epsfig{file=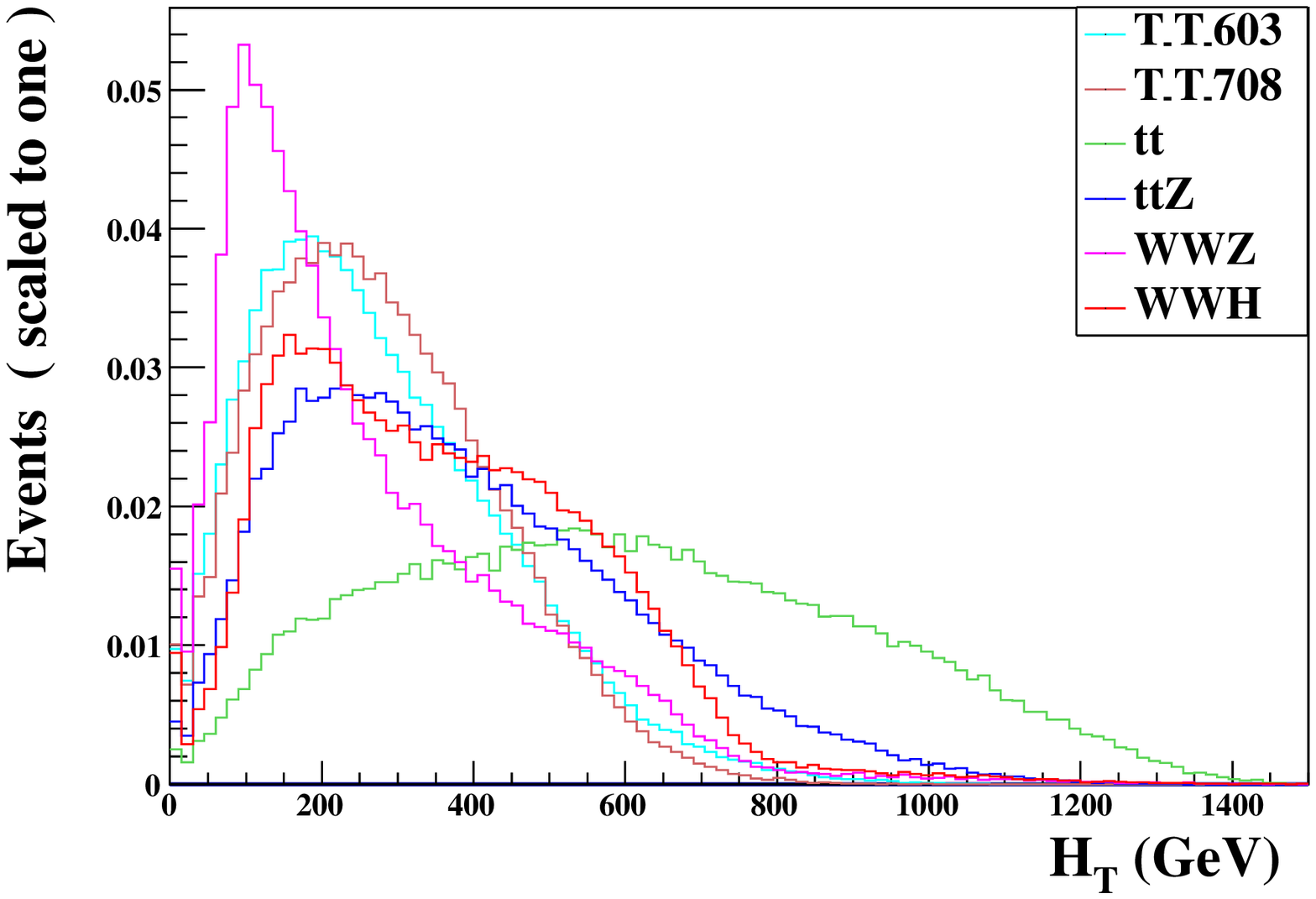}} \caption{Normalized
distributions of $\eta_{l_{1}}$, $\eta_{b_{1}}$, $p_{T}^{l1}$,
$\Delta R_{l_{1}b_{1}}$, $E(b_{1}l_{1})$, $H_{T}$ in the signal and
backgrounds for the two signal benchmark points at $\sqrt{s}=1.5$
TeV. }\label{tmtm}
\end{center}
\end{figure}
Take into consideration the constraints on the top partner mass from
current measurements, we take $f=700$GeV, $R=1$ (correspond to
$m_{T_{-}}=708$ GeV) and $f=700$GeV, $R=1.5$ (correspond to
$m_{T_{-}}=603$ GeV) for two benchmark points in the following
calculations. In order to reduce the background contribution and
enhance the signal contribution, some cuts of kinematic
distributions are needed. In Fig.\ref{tmtm}, we show the normalized
distributions of transverse momentum $p_{T}^{l_{1}}$, the
pseudorapidity $\eta_{l_{1}}$, $\eta_{b_{1}}$, the separation
$\Delta R(l_{1},b_{1})$ between $l_{1}$ and $b_{1}$, the energy
$E(b_{1}l_{1})(=E(b_{1})+E(l_{1}))$ and the total transverse energy
$H_{T}$.

Since the dominant background arises from $t\bar{t}$, the cuts that
are chosen to suppress the backgrounds should centered around the
$t\bar{t}$ background. Firstly, we can apply the cuts of general
kinematic distributions, such as $p_{T}^{l_{1}}$, $\eta_{l_{1}}$,
$\eta_{b_{1}}$ to suppress the backgrounds. For the the $\Delta
R(l_{1},b_{1})$ distribution, there are two peaks in the $t\bar{t}$,
$t\bar{t}Z$ backgrounds and one peak in the $WWZ$, $WWH$
backgrounds, we can use the deviation between the signal peak and
background peak to suppress the backgrounds. Then, in view of the
energy $E(b_{1}l_{1})$ distribution, we can also use the deviation
between the signal peak and background peak to reduce the
backgrounds. After that, the $H_{T}$ distribution of the signal can
be utilized to remove the $t\bar{t}$ background effectively.
According to the above analysis, events are selected to satisfy the
following cuts:
\begin{eqnarray}
&&\textrm{Cut-1}: p_{T}(l_{1})>50\textrm{GeV};\nonumber\\
&&\textrm{Cut-2}: |\eta(l_{1})|<1;|\eta(b_{1})|<1;\nonumber\\
&&\textrm{Cut-3}:  \Delta R(l_{1},b_{1})<2.5;\nonumber\\
&&\textrm{Cut-4}:~E(b_{1}l_{1})<400\textrm{GeV}; \nonumber\\
&&\textrm{Cut-5}: H_{T}<400\textrm{GeV};\nonumber
\end{eqnarray}

\begin{table}[ht]
\caption{Cut flow of the cross sections for the signal(S) and the
backgrounds(B) for the two signal benchmark points (P1: $f=700$ GeV,
$R=1$) and (P2: $f=700$ GeV, $R=1.5$) at $\sqrt{s}$=1.5TeV.
\label{tab2}}
\bigskip
\begin{tabular}{|c|c|c|c|c|c|c|c|c|c|}
\hline
     \multirow{2}{*}{Cuts}& \multicolumn{2}{c|}{S($\times 10^{-3}$fb)} &\multicolumn{4}{c|}{B($\times 10^{-3}$fb)}&\multicolumn{2}{c|}{S/B}\\
\cline{2-9}
    & $T_{-}\bar{T}_{-}$(P1) &$T_{-}\bar{T}_{-}$(P2) & $t\bar{t}$&  $t\bar{t}Z$ &$WWZ$&$WWH$&P1 &P2 \\
\hline
    No cut&184&119&3485&32&367&100&0.046&0.03 \\
\hline    Cut-1&139.8&94.0&2011&20.8&283&104&0.058&0.039\\
\hline
   Cut-2&81.1&54.9&929.6&9.6&59.8&41.1&0.078&0.053\\
\hline
   Cut-3&62.4&45.6&334.7&5.6&15.6&11.5&0.17&0.12 \\
\hline
   Cut-4&48.7&36.5&120.1&3.4&3.0&2.2&0.38&0.28 \\
\hline
Cut-5&44.8&33.6&34.8&2.4&2.6&1.5&1.08&0.81 \\
\hline
\end{tabular}
\end{table}

\begin{figure}[htbp]
\begin{center}
\scalebox{0.35}{\epsfig{file=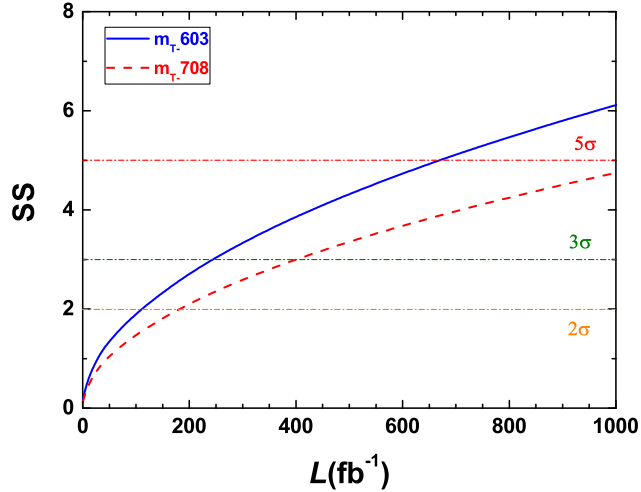}}\vspace{-0.5cm}\hspace{-0.cm}
\caption{The statistical significance depending on integrated
luminosity for $\sqrt{s}$=1.5TeV.}\label{ssl}
\end{center}
\end{figure}

For easy reading, we summarize the cut-flow cross sections of the
signal and backgrounds for c.m. energy $\sqrt{s}$=1.5 TeV in Table
\ref{tab2}. To estimate the observability quantitatively, the
Statistical Significance ($SS$) is calculated after final cut by
using Poisson formula\cite{Poisson}
\begin{eqnarray}
SS=\sqrt{2L\left [ (S+B)\ln\left(1+\frac{S}{B}\right )-S\right ]},
\end{eqnarray}
where $S$ and $B$ are the signal and background cross sections and
$L$ is the integrated luminosity. The results for the $SS$ values
depending on the integrated luminosity for $\sqrt{s}$=1.5TeV are
shown in Fig.\ref{ssl}. It is clear from Fig.\ref{ssl} that we can
obtain the $2\sigma$ significance at a luminosity of 110(200)
fb$^{-1}$ and $3\sigma$ significance at a luminosity of 250(400)
fb$^{-1}$ for $m_{T_{-}}$=603(708)GeV.

\section{Conclusions}
In this paper, we discuss the top partner production at future
$e^{+}e^{-}$ collider in the LHT model. We first consider the
constraints on the top partner masses from the current measurements,
then calculate the cross sections of various top partner production
processes, which includes $e^{+}e^{-}\rightarrow T_{+}\bar{T}_{+}$,
$e^{+}e^{-}\rightarrow T_{-}\bar{T}_{-}$, $e^{+}e^{-}\rightarrow
t\bar{T}_{+}$ and $e^{+}e^{-}\rightarrow T_{+}\bar{T}_{+}H$,
$e^{+}e^{-}\rightarrow T_{-}\bar{T}_{-}H$ and $e^{+}e^{-}\rightarrow
t\bar{T}_{+}H$. Next, we investigate the observability of the T-odd
top partner pair production through the process
$e^{+}e^{-}\rightarrow T_{-}\bar{T}_{-}\rightarrow
t\bar{t}A_{H}A_{H}$ with the di-lepton decay of the top quark pair
for $\sqrt{s}$=1.5TeV. We display the signal significance depending
on the integrated luminosity and find that the $2\sigma$
significance can be obtained at a luminosity of 110(200) fb$^{-1}$
for $m_{T_{-}}$=603(708)GeV, which is promising at the future high
energy $e^{+}e^{-}$ collider with high luminosity.

\section*{Conflicts of interest}
The authors declare that there is no conflict of interest regarding
the publication of this paper.

\section*{Acknowledgement}
This work is supported by the National Natural Science Foundation of
China (NNSFC) under grants No. 11405047, No.11404099, by the Startup
Foundation for Doctors of Henan Normal University under Grant No.
qd15207.

\end{document}